# Dominant itinerant ferromagnetism in $Eu_{0.5}Sr_{0.5}CoO_3$: evidences from both critical analysis and Rhodes-Wohlfarth's criterion

Renwen Li[a,b,*], Feng Li[a], Jun Fang[a], Wei Tong[a], Changjin Zhang[a], Li Pi[a], and Yuheng Zhang[a,*]

[a]*High Magnetic Field Laboratory, Chinese Academy of Sciences, Hefei 230031, People's Republic of China*

[b]*Department of Physics and Electronic Engineering, Hefei Normal University, Hefei 230061, People's Republic of China*

**Abstract**

The critical behaviors of ferromagnet $Eu_{0.5}Sr_{0.5}CoO_3$ arround $T_C$=140.5K have been comprehensively investigated by analyzing a series of isothermal magnetization M(H) curves. Both Modified Arrott plot and Kouvel-Fisher methods give nearly the same critical exponents, which scale nicely the M(H) curves into two different branches below and above $T_C$. The exponents $\gamma$=1.044 and $\delta$=3.06 demonstrate the relevance of mean-field characters for this material. The conclusion of mean-field behavior proves a dominant itinerant ferromagnetism (FM) due to a long range exchange interaction in the system. Meanwhile, by using Rhodes-Wohlfarth's criterion [P. Rhodes and E. P. Wohlfarth, Proc. R. Soc. Lond. A **273**, 247 (1963)], it is further confirmed that the itinerant FM dominates in the system.

*Key words:*

$Eu_{0.5}Sr_{0.5}CoO_3$

Critical exponent

Itinerant ferromagnetism

Rhodes-Wohlfarth's criterion

## 1. Introduction

    Perovskite compounds have attracted considerable attention in the past few decades due to their potential applications, such as magnetic storage and magnetic

---
* Corresponding author.
  E-mail adresses: lirw@hmfl.ac.cn (R. Li), zhangyh@ustc.edu.cn (Y. Zhang).

refrigeration. Among them, perovskite cobaltites $Ln_{1-x}A_xCoO_3$ (Ln=rare earth, A=alkaline earth) have been focused in recent years [1-25]. Since the energy difference between $t_{2g}$ and $e_g$ levels is very small, the spin state of the Co ions can be in low-spin, intermediate-spin, or high-spin state with the application of external stimuli from temperature, compositional doping, magnetic field, or pressure[1-11]. $LnCoO_3$ has a low-spin $Co^{3+}$ at ground state as a nonmagnetic insulator. By substituting Ln with alkaline earth ions like Ca or Sr, hole-rich ferromagnetic (FM) regions and hole-poor antiferromagnetic (AFM) regions are generated and coexist in a microscopic scale, indicating $Ln_{1-x}Sr_xCoO_3$ is not a homogeneous ferromagnet [1-29]. This microscopic inhomogeneity may cause the complex exchange interaction, and thereby, the paramagnetic (PM) to FM phase transition.

It is known that an analysis of the critical behaviors on a FM transition is an effective way to understand the order of the transition and the intrinsic nature of a ferromagnet. An early study on the critical behavior in $La_{1-x}Sr_xCoO_3$ ($0.2 \leq x \leq 0.3$) by Mira *et al.* [30] has shown that the critical exponent γ is in accord with Heisenberg model while β is mean-field-like. A detailed analysis of the critical behavior in single crystal $La_{0.67}Sr_{0.33}CoO_3$ has shown that all the exponents values match well with the three-dimensional (3D) Heisenberg values with nearest-neighbor interaction [31]. In half-doped $La_{0.5}Sr_{0.5}CoO_3$, a previous study [32] of critical exponents has suggested that γ value is close to 3D Ising value whereas the exponent δ approximately equals to mean-field one. Differently, Mukherjee *et al.* [33] have suggested that all the values of critical exponents for the same composition correspond to Heisenberg class. Recently, Khan *et al.* [25] examined the critical behavior in $La_{1-x}Sr_xCoO_3$ (*x*=0.21 and 0.25) single crystals and found a deviation of the critical exponents from Heisenberg towards mean-field. Hence, in order to understand the nature of the FM transition in this system, it is important to investigate the critical exponents in various cobaltite compounds. To our knowledge, the studies have been performed largely on lanthanum-based cobaltites, whereas there are few reports on similar compositions of other rare earth element (such as Nd, Sm, and Eu, *et al.*) based cobaltites.

In this paper, we present a detailed investigation of the critical behaviors in ferromagnet $Eu_{0.5}Sr_{0.5}CoO_3$. It is found that the $Eu_{0.5}Sr_{0.5}CoO_3$ is in good accordance with the mean-field model, in which the long-range exchange interaction dominates. This is distinctly different from that of $La_{0.5}Sr_{0.5}CoO_3$, where the short range Heisenberg model works. Based on the Rhodes-Wohlfarth's criterion [34], it is further proved that the itinerant FM dominates in $Eu_{0.5}Sr_{0.5}CoO_3$. The dominant itinerant FM interaction would enlarge the range of the exchange interaction, which may be responsible for the mean-field behavior.

## 2. Experiment

Polycrystalline sample of $Eu_{0.5}Sr_{0.5}CoO_3$ was synthesized using the conventional solid-state reaction method as described elsewhere [35]. It was confirmed in single phase with cubic $Pm\bar{3}m$ space group by the powder x-ray diffraction (XRD), which is consistent with reference code 00-053-0113 for $Eu_{0.5}Sr_{0.5}CoO_3$ from the previous report [36]. For $Ln_{0.5}Sr_{0.5}CoO_3$ series, the reducing size of the Ln cation should lead to stronger distortion from cubic to orthorhombic. As shown in the Table 1 from the reference [36], it is true for Pr (monoclinic $P2_1/n$), Nd (orthorhombic *Pnma*), and Sm (orthorhombic *Pnma*), while for Eu, the symmetry (space group) is still cubic ($Pm\bar{3}m$). The magnetic measurements were performed on a commercial superconducting quantum interference device (SQUID) magnetometer (Quantum Design MPMS). The sample for magnetic measurements was made in a slender ellipsoid shape. All the magnetic data were collected with the field applied parallel to the longest axis to minimize the demagnetization effect. The demagnetization factor D can be determined from the slope of the low-field $M(H_a)$ data, yielding the internal field $H_i=H_a-4\pi DM$. We compare the M(H) and Arrott plots for the raw data and the amended data by subtracting the demagnetizing field contribution, and find the demagnetizing field has little effect on the analysis below. The isothermal magnetizations were performed at 1 or 2K interval over the temperature range from 125 to 150K with field sweeping from 50 to 45kOe. Before each magnetization measurement, the sample was demagnetized and warmed to room temperature (well above the Curie temperature $T_C$) for enough time and then cooled down to the target temperature so that all curves were initial magnetizing and started from the same magnetization state.

### 3. Results and discussions

Figure 1(a) shows the temperature variation of field cooled (FC) magnetization M(T) for $Eu_{0.5}Sr_{0.5}CoO_3$ under magnetic fields of 0.01, 0.1 and 0.5T. It shows a PM-FM phase transition at $T_C \approx 140K$. The $T_C$ is estimated from the temperature dependence of the differential quotient of magnetization dM/dT, as shown in Fig. 1(b), which is consistent with previous reports [36,37]. And the value is approximately in-line with $T_C(x)$ of $Eu_{1-x}Sr_xCoO_3$ [37]. The dM/dT vs. T curves exhibit symmetric peaks with full width at half maximum ($T_{FWHM}$) of about 12K at low magnetic field of 0.01T. As shown in the left inset of Fig. 1(b), $T_{FWHM}$ increases almost linearly with increasing applied field at a rate of about 28.6±0.7K/T. According to N. Khan *et al.* [25], the symmetric peak implies the high quality and the crystalline nature of the samples. As shown in Fig. 1 (b) and its right inset, the minimum of dM/dT locates at $T_C$, but the $T_C$ value is enhanced greatly with field at a rate of ~14.2±1.5K/T.

For a second order magnetic phase transition, the critical behaviors can be characterized using a set of critical exponents, β (associated with the spontaneous magnetization $M_S$), γ (associated with the initial susceptibility $\chi_0$) and δ (associated with critical magnetization isotherm). The definitions of the exponents from magnetization can be described as [38]:

$$M_S(T) = M_0(-\varepsilon)^\beta, \varepsilon < 0, T < T_C \quad (1)$$

$$\chi_0(T)^{-1} = (h_0/M_0)\varepsilon^\gamma, \varepsilon > 0, T > T_C \quad (2)$$

$$M = DH^{1/\delta}, \varepsilon = 0, T = T_C \quad (3)$$

where $\varepsilon=(T-T_C)/T_C$ denotes the reduced temperature, $h_0/M_0$ and D are the critical amplitudes.

The Arrott plot ($M^2$ vs. H/M plot, i.e. β=0.5 and γ=1) has been shown in our previous paper [35]. According to the criterion proposed by Banerjee [39,40], the positive slopes of the $M^2$ vs. H/M plot suggest a second order transition in this system. Based on the mean-field theory, the regular Arrott plot near the transition should be a set of parallel lines in the high field regimes, and the line at $T_C$ should pass through the origin [41]. In the Arrott plot for the present case, all curves show nearly parallel lines in the high field regime, and the linear extrapolation from high field regime to the intercepts yields the values of $M_S(T,0)$ for $T<T_C$ and $\chi_0^{-1}(T,0)$ for $T>T_C$, respectively. Using Eqs. (1) and (2), the temperature variation of $M_S(T,0)$ and $\chi_0^{-1}(T,0)$ are plotted and fitted. The fitting results produce two new critical exponents as β and γ, which are used to make the new modified Arrott plot. The procedure is performed repeatedly until the values of β and γ do not change. The final $M_S(T)$ and $\chi_0^{-1}(T)$ curves are shown in Fig. 2, which gives two sets of critical exponents of β=0.412±0.013 with $T_C$=140.49±0.12K and γ=1.044±0.019 with $T_C$=140.67±0.08K.

Kouvel and Fisher [42] have also put forward an effective method (namely KF method) which can deduce the critical exponents $T_C$, β and γ:

$$\frac{M_S(T)}{dM_S(T)/dT} = \frac{T-T_C}{\beta} \tag{4}$$

$$\frac{\chi_0^{-1}(T)}{d\chi_0^{-1}(T)/dT} = \frac{T-T_C}{\gamma} \tag{5}$$

The KF method suggests that the temperature dependences of $M_S(T)/[dM_S(T)/dT]$ and $\chi_0^{-1}(T)/[d\chi_0^{-1}(T)/dT]$ should give straight lines with slopes 1/β and 1/γ, and intercept of $T_C$ on T axis, respectively. Using the values of $M_S(T,0)$ and $\chi_0^{-1}(T,0)$ determined from Arrott plot, the temperature variation of $M_S(T)/[dM_S(T)/dT]$ and $\chi_0^{-1}(T)/[d\chi_0^{-1}(T)/dT]$ curves are plotted in Fig. 3. The KF linear fitting gives the new exponents as β=0.415±0.002 with $T_C$=140.53±0.05K and γ=1.044±0.004 with $T_C$=140.67±0.02K. These critical exponents obtained from the KF method agree well with those obtained from the modified Arrott plot in Fig. 2. The third exponent δ can be obtained by using the Widom scaling relation as [43]:

$$\delta = 1+\gamma/\beta \tag{6}$$

It gives δ=3.53 from Fig. 2 and 3.51 from Fig. 3. Compared with the values in 3D-Heisenberg, 3D-Ising and Tricritical mean-field models, as shown in Table I, both approach to the value (3.0) in mean-field model.

The deduced exponents can be tested with the prediction of the scaling hypothesis as [39]:

$$M(H,\varepsilon) = \varepsilon^{\beta} f_{\pm}(H/\varepsilon^{\beta+\gamma}) \quad (7)$$

where $f_{\pm}$ are regular functions with $f_{+}$ for T>$T_C$ and $f_{-}$ for T<$T_C$. The scaling relation predicts that $M(H,\varepsilon)\varepsilon^{-\beta}$ vs. $H\varepsilon^{-(\beta+\gamma)}$ should yield two universally different curves, one for T>$T_C$ and the other for T<$T_C$. Taking the values of β, γ and $T_C$ from the KF method in Fig. 3, the isothermal magnetizations around $T_C$~140K are plotted on a log-log scale in Fig. 4. All the magnetization data fall into two curves, one for T>$T_C$ and the other for T<$T_C$. This result indicates that the obtained critical exponents are intrinsic and in agreement with the scaling hypothesis.

Figure 5 shows the M(H) curves on a log-log scale near $T_C$. Based on Eq. (3) and the obtained $T_C$ values above, the M(H) curve at 140K was selected as the critical isothermal magnetization. The straight line with a slope of 1/δ in high field region gives δ=3.06±0.03. This is in excellent accord with the value of δ=3.0 in mean-field model.

For the ferromagnet $Eu_{0.5}Sr_{0.5}CoO_3$, it is found that γ=1.044 and δ=3.06 approach mostly to the values in mean-field model (γ=1.0 and δ=3.0) except that β=0.415 is a little smaller than that in mean-field model (0.5) but close to that in 3D-Heisenberg model (β=0.365). Actually, the value of β locates between the 3D-Heisenberg model and the mean-field model. Figure 6(a-c) show the modified Arrott plots with three different models as: the 3D-Heisenberg model (β=0.365 and γ=1.386), 3D-Ising model (β=0.325 and γ=1.24) and Tricritical mean-field model (β=0.25 and γ=1.0). It is found that all the three models yield quasi-straight parallel lines in the high field regime, similar in mean-field model reported previously [35]. In order to distinguish more clearly and determine the most suitable model, the temperature variation of relative slope RS(T) is plotted in Fig. 7. Here RS(T) is defined as S(T)/S($T_C$), S(T) is the slope of the quasi-straight line in the high field region at T. Since 140K is closest to $T_C$ value, we select S(140K) as S($T_C$). In the most ideal model, all RSs should be equal to 1 because the ideal modified Arrott plot is a series of parallel lines. As for $Eu_{0.5}Sr_{0.5}CoO_3$, the RSs in mean-field model approach to the ideal value 1 best, while in the other three models the RSs apparently deviate from 1. This fact suggests that the mean-field model is the most suitable one to describe the critical phenomena for $Eu_{0.5}Sr_{0.5}CoO_3$.

A renormalization group analysis has suggested that the values of the critical exponents depend on and reflect the range of the exchange interaction J(r) in a d dimension system with the form as [44]:

$$J(r) \sim r^{-(d+\sigma)} \tag{8}$$

where r and σ are the distance and the range of the exchange interaction. For a three dimension isotropic system (d=3), it complies with the 3D-Heisenberg model (β=0.365, γ=1.386 and δ=4.797) only if σ≥2, i.e., if J(r) decreases with "short-range" distance faster than $r^{-5}$. Whereas if σ≤3/2 the mean-field model (β=0.5, γ=1.0 and δ=3.0) is valid, which indicates J(r) decreases with "long-range" distance slower than $r^{-4.5}$. In the intermediate range 3/2≤σ≤2, J(r) decays as $r^{-(3+\sigma)}$, the system belongs to different classes with exponents taking intermediate values depending on σ value [44]. A renormalization group analysis [44,45] has suggested that the exponent of γ and the range of σ satisfy a mathematic relation as:

$$\gamma = 1 + \frac{4}{d}\left(\frac{n+2}{n+8}\right)\Delta\sigma + \frac{8(n+2)(n-4)}{d^2(n+8)^2}\left[1 + \frac{2G(\frac{1}{2}d)(7n+20)}{(n-4)(n+8)}\right]\Delta\sigma^2 \tag{9}$$

where $\Delta\sigma = \sigma - \frac{1}{2}d$ and $G(\frac{1}{2}d) = 3 - \frac{1}{4}(\frac{1}{2}d)^2$ for the system with dimensionality of lattice (d) and spin (n). In the present case, based on γ=1.044±0.004 one can calculate σ=1.568±0.001, which indicates J(r) decays as $r^{-4.57}$. Obviously, J(r) approaches to the mean-field model, implying a "long range" spin interaction.

It is clear from above discussions that FM interaction in $Eu_{0.5}Sr_{0.5}CoO_3$ does not comply with the 3D short-range interaction model. Generally, the intermediate range of the interaction could be put down to either the long-range dipolar interaction [46-48] or the long-range interaction between spins. The dipolar-FM can be excluded in the Eu$_{0.5}$Sr$_{0.5}$CoO$_3$ compound due to the apparent deviation between the exponents listed in Table 1 and that for dipolar-FM model [49]. So it is important to make clear the nature of FM interaction in this system.

Generally, for transition ions in Ln$_{0.5}$Sr$_{0.5}$CoO$_3$, the orbital angular momentum is assumed to be fully quenched by the crystal field effect, which leaves a spin only moment. According to Rhodes-Wohlfarth's criterion [34], one can distinguish whether the nature of the ferromagnetism is localized or itinerant based on the ratio of $q_C/q_S$, where $q_C$ and $q_S$ are the values for the magneton numbers calculated from the Curie-Weiss constant above T$_C$ and the saturation magnetization M$_S$ in low temperature, respectively. For the localized FM, M$_S$ equals to the fully aligned spin moment, giving $q_C/q_S$ =1. While for the itinerant FM, M$_S$ is less than the fully aligned spin moment, giving $q_C/q_S$ >1.

The $q_C$ can be determined from the effective PM moment ($p_{eff} = g\mu_B\sqrt{S(S+1)}$, where S=$q_C/2$ is the effective spin numbers per atom, $\mu_B$ is the Bohr magneton, and $g\approx2$ is the Lande $g$-factor for transition metals). For a reasonable analysis, the magnetic contribution from the Eu$^{3+}$ should be eliminated before using the Rhodes-Wohlfarth's criterion. We assume that the Eu$^{3+}$ ions are free and paramagnetic. The effective moment for free Eu$^{3+}$ ion is 3.32 $\mu_B$ taken from Van Vleck [50]. The right axis of Fig. 1(a) shows the Curie-Weiss linear behavior in high temperature PM regime and yields Curie constant C=2.1 $emu\cdot K/mol$ Oe. This gives $p_{eff}$ =4.1 $\mu_B$, S=1.61 and $q_C$ =3.22 including Eu$^{3+}$ contribution, which are all higher than those in La$_{0.5}$Sr$_{0.5}$CoO$_3$ (C=1.85 $emu\cdot K/mol$ Oe, $p_{eff}$ =3.85 $\mu_B$, S=1.49 and $q_C$ =2.98) [15,16,32,51], possibly due to the contribution from the high temperature PM response from Eu$^{3+}$. After subtracting the contribution from Eu$^{3+}$, $(p_{eff})_{Co} = \sqrt{(p_{eff}^2)_{Eu_{0.5}Sr_{0.5}CoO_3} - 0.5(p_{eff}^2)_{Eu}}$, the effective moment turns out to be 3.36 $\mu_B$, giving S=1.253 and $q_C$=2.506. As shown the M(H) curves from the reference [35], compared with the large FM contribution from Co ions, the magnetic contribution from $Eu^{3+}$ is very small and can be neglected. So we take the magnetic moment value at 5K in 1T field as the saturation moment $M_S = q_s\mu_B$, where the magnetization increases steeply and just starts to saturate. The $q_S$ value for $Eu_{0.5}Sr_{0.5}CoO_3$ is estimated to be 0.86, which is much smaller than 1.79 in its counterpart $La_{0.5}Sr_{0.5}CoO_3$ [16,32,33,52]. This indicates that $Eu_{0.5}Sr_{0.5}CoO_3$ has a different magnetic exchange interaction from $La_{0.5}Sr_{0.5}CoO_3$. The ratio of $q_C/q_S$ for $Eu_{0.5}Sr_{0.5}CoO_3$ can be determined to be 2.91 which is much higher than 1, indicating itinerant FM behavior in $Eu_{0.5}Sr_{0.5}CoO_3$. While for $La_{0.5}Sr_{0.5}CoO_3$ the ratio of $q_C/q_S$ is only 1.66 nearly approaching to 1. Goodenough [53] has suggested a coexistence of localized and itinerant d electrons in $LaCoO_3$ and $La_{0.5}Sr_{0.5}CoO_3$ systems, which has been experimentally established by Mössbauer study and theoretical calculation [15,54,55]. Therefore, it is reasonable to conclude that, in $(Ln,Sr)CoO_3$ system, there exists a competition between localized and itinerant FM. The localized FM may dominate in $La_{0.5}Sr_{0.5}CoO_3$ leading to a short-range Heisenberg spin interaction, while the itinerant FM dominates in $Eu_{0.5}Sr_{0.5}CoO_3$ resulting in a long-range mean-field spin interaction. The dominance of itinerant FM in $Eu_{0.5}Sr_{0.5}CoO_3$ leads that the saturation magnetization is weaker than that in $La_{0.5}Sr_{0.5}CoO_3$ where localized FM dominates. The dominant itinerant FM due to the long range of the exchange interaction may lead that J(r) approaches towards mean-field model. Furthermore, the field dependence of the magnetization M(H) for $Eu_{0.5}Sr_{0.5}CoO_3$ is weaker than that for $La_{0.5}Sr_{0.5}CoO_3$, while it is still rather strong. This also demonstrates that, the localized FM and itinerant FM coexist in the system but the itinerant FM dominates.

**4. Conclusion**

In summary, the critical behaviors of $Eu_{0.5}Sr_{0.5}CoO_3$ have been comprehensively studied near the critical point $T_C$ using the modified Arrott plot, Kouvel-Fisher method and critical isotherm analysis. The critical exponents are calculated as $\beta=0.415\pm0.002$, $\gamma=1.044\pm0.004$ and $\delta=3.06\pm0.03$ as well as $T_C\approx140K$. The results indicate that the magnetic behaviors of $Eu_{0.5}Sr_{0.5}CoO_3$ approach to the mean-field model rather than the Heisenberg one. This proves that the itinerant FM dominates in the system due to a long range exchange interaction, which is further tested by the Rhodes-Wohlfarth's criterion.

## Acknowledgements

We wish to acknowledge financial supports from the State Key Project of Fundamental Research of China (No. 2010CB923403), the National Natural Science Foundation of China (No. 11004194), the Natural Science Foundation of the Anhui Higher Education Institutions of China (No. KJ2011Z318), Hefei Normal University Research Funding (No. 2012kj02), China Postdoctoral Science Foundation funded project (No. 2012M511428), and Anhui Provincial Natural Science Foundation (No. 1208085QA17).

**Figure captions**

**Fig. 1.** The temperature dependence of (a) magnetization (left axis), H/M (right axis, the solid line is fit to the Curie-Weiss law) and (b) dM/dT in FC mode under 0.01T, 0.1T and 0.5T field for $Eu_{0.5}Sr_{0.5}CoO_3$. The insets of (b) show the field dependence of $T_{FWHM}$ and $T_C$.

**Fig. 2.** Temperature dependence of the spontaneous magnetization $M_S(T, 0)$ and the inverse initial susceptibility $\chi_0^{-1}(T)$ along with the fitting curves (solid lines) with the help of power law due to Eqs. (1) and (2).

**Fig. 3.** KF plot for $M_S(T)$ and $\chi_0^{-1}(T)$, solid lines are the linear fitting of the data.

**Fig. 4.** Scaling plot below and above $T_C$ using exponents determined from the KF method (only several typical curves are shown).

**Fig. 5.** Isothermal M(H) plots in a log-log scale around $T_C$, the solid line is the linear fitting following Eq. (3) at 140 K.

**Fig. 6.** Modified Arrott plots: (a) 3D-Heisenberg model ($\beta=0.365$, $\gamma=1.386$), (b) 3D-Ising model ($\beta=0.325$, $\gamma=1.24$) and (c) Tricritical mean-field model ($\beta=0.25$, $\gamma=1.0$).

**Fig. 7.** The temperature variation of the relative slope RS [RS≡S(T)/S($T_C$)].

**Tables captions:**

**Table 1:** Derived critical exponents for $Eu_{0.5}Sr_{0.5}CoO_3$ with different theoretical models and other analogous compounds reported in literatures. Abbreviations: MAP=modified Arrott plot; KFM=Kouvel-Fisher method; WSR=Widom scaling relation.

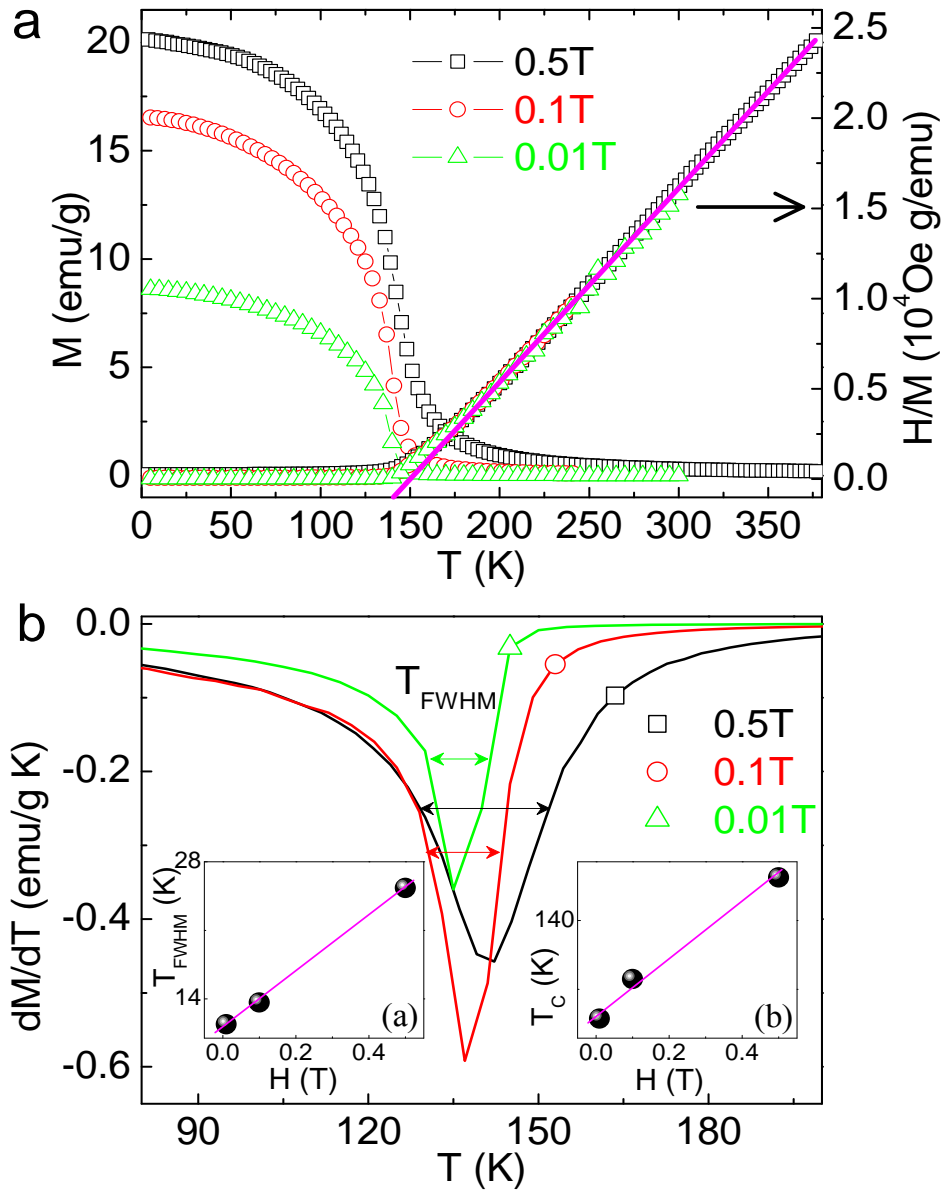

**Fig. 1.** The temperature dependence of (a) magnetization (left axis), H/M (right axis, the solid line is fit to the Curie-Weiss law) and (b) dM/dT in FC mode under 0.01T, 0.1T and 0.5T field for $Eu_{0.5}Sr_{0.5}CoO_3$. The insets of (b) show the field dependence of $T_{FWHM}$ and $T_C$.

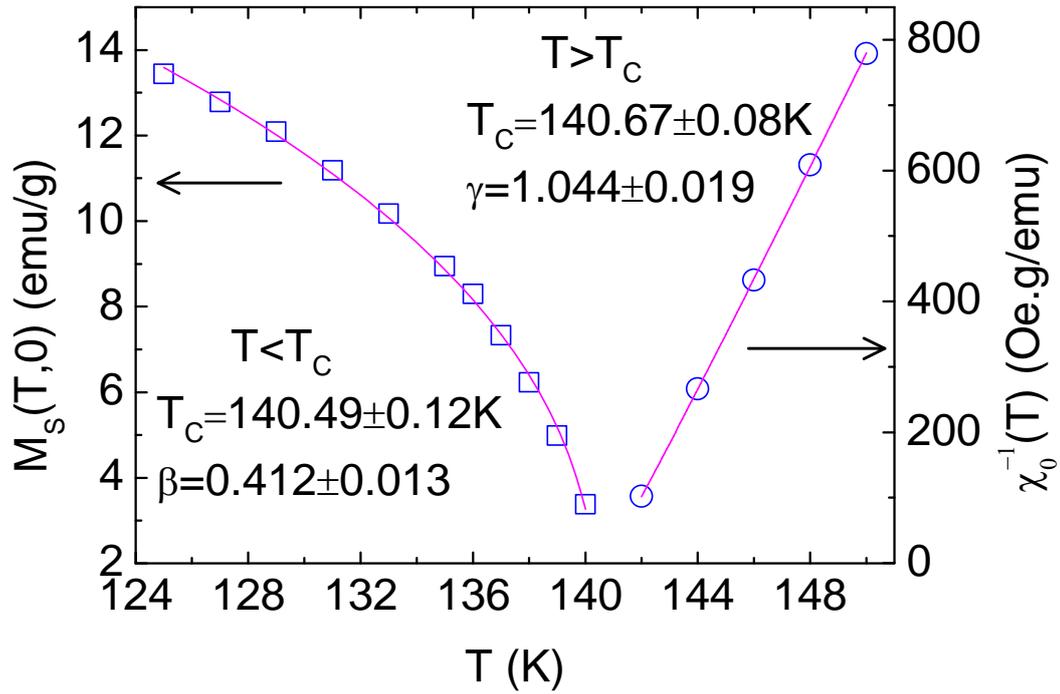

**Fig. 2.** Temperature dependence of the spontaneous magnetization $M_S(T, 0)$ and the inverse initial susceptibility $\chi_0^{-1}(T)$ along with the fitting curves (solid lines) with the help of power law due to Eqs. (1) and (2).

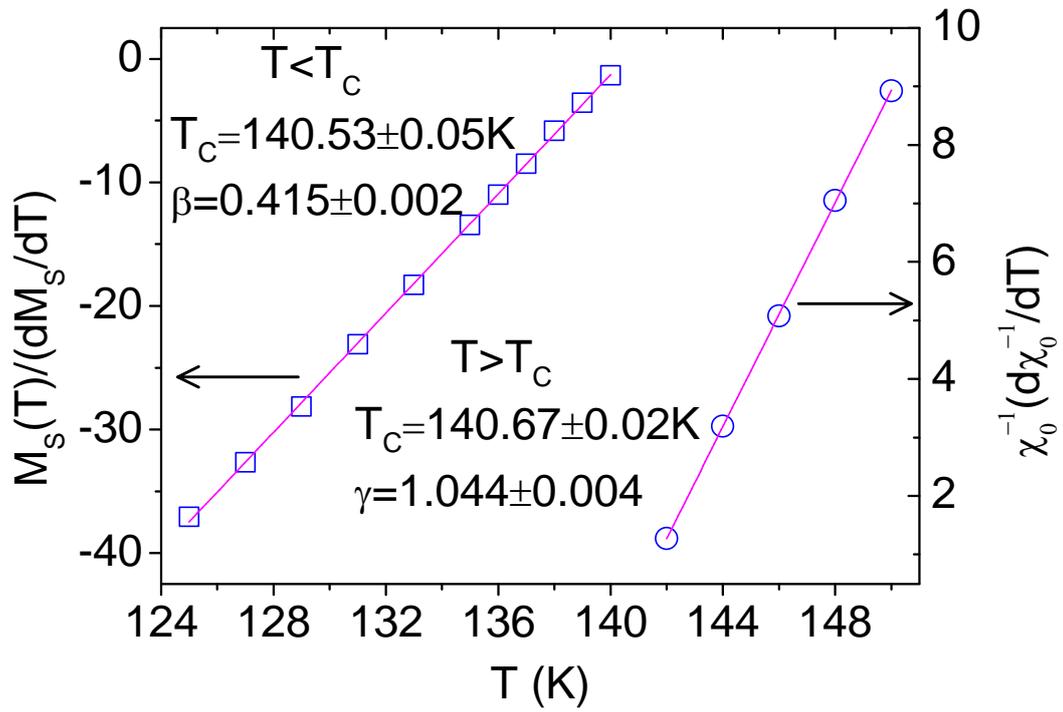

**Fig. 3.** KF plot for $M_S(T)$ and $\chi_0^{-1}(T)$, solid lines are the linear fitting of the data.

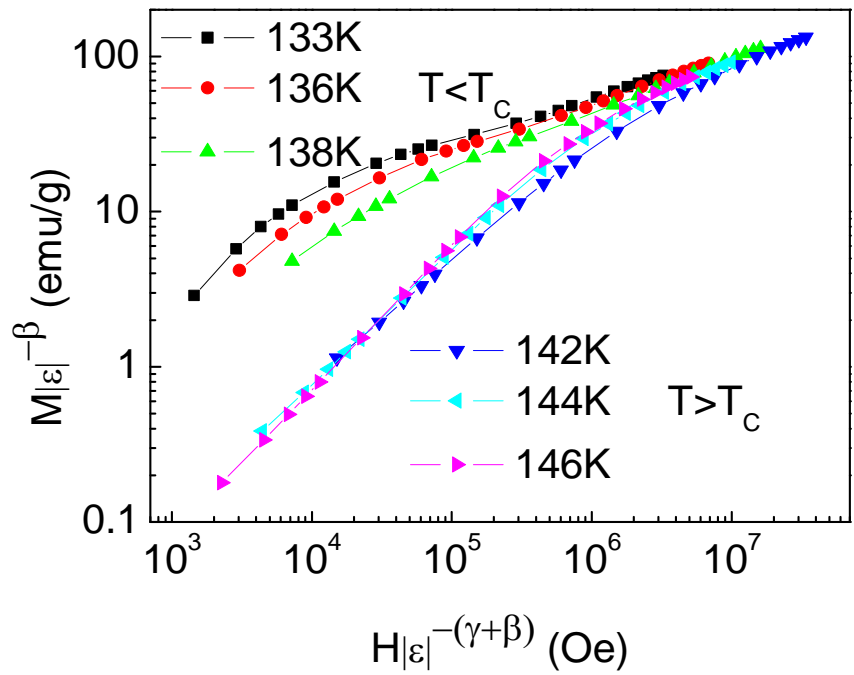

**Fig. 4.** Scaling plot below and above $T_C$ using exponents determined from the KF method (only several typical curves are shown).

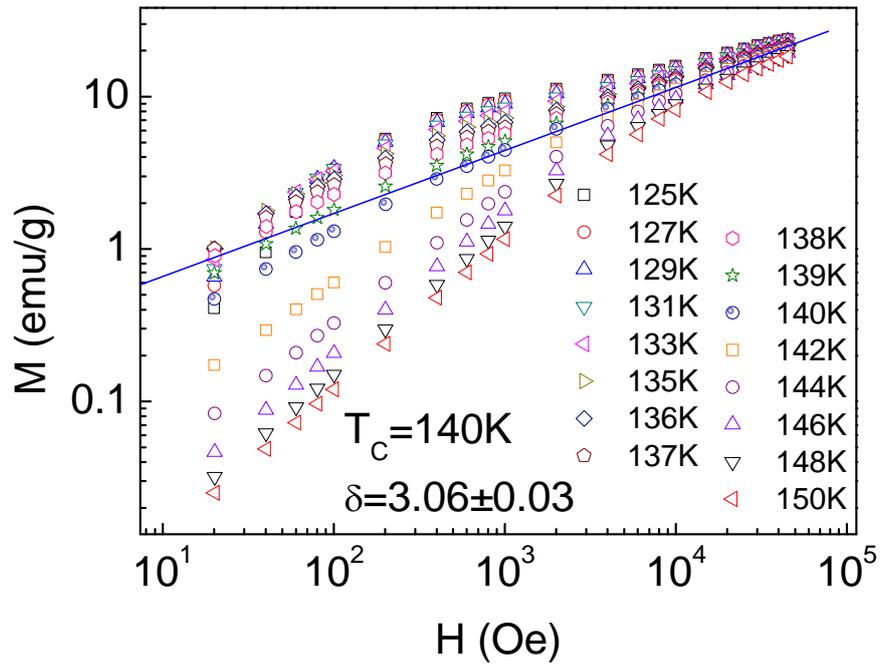

**Fig. 5.** Isothermal M(H) plots in a log-log scale around $T_C$, the solid line is the linear fitting following Eq. (3) at 140 K.

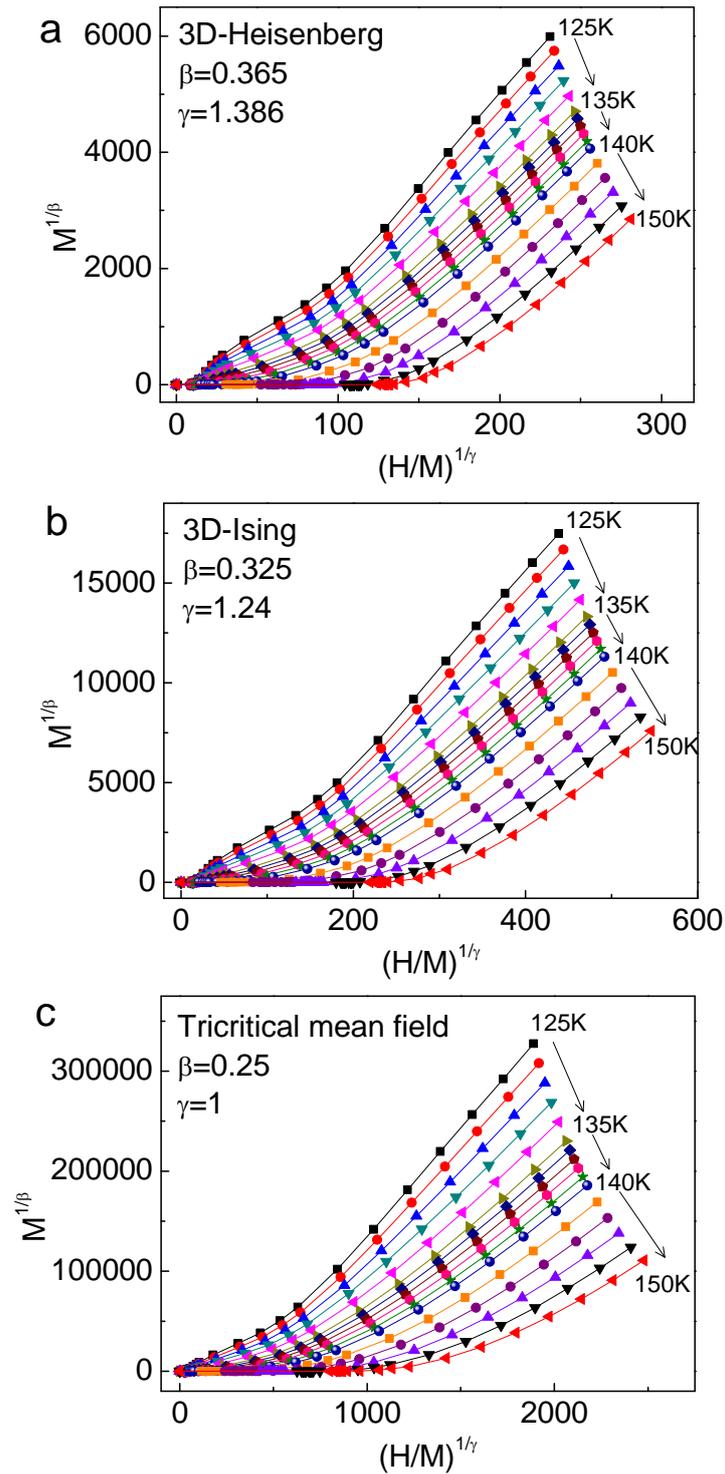

**Fig. 6.** Modified Arrott plots: (a) 3D-Heisenberg model (β=0.365, γ=1.386), (b) 3D-Ising model (β=0.325, γ=1.24) and (c) Tricritical mean-field model (β=0.25, γ=1.0).

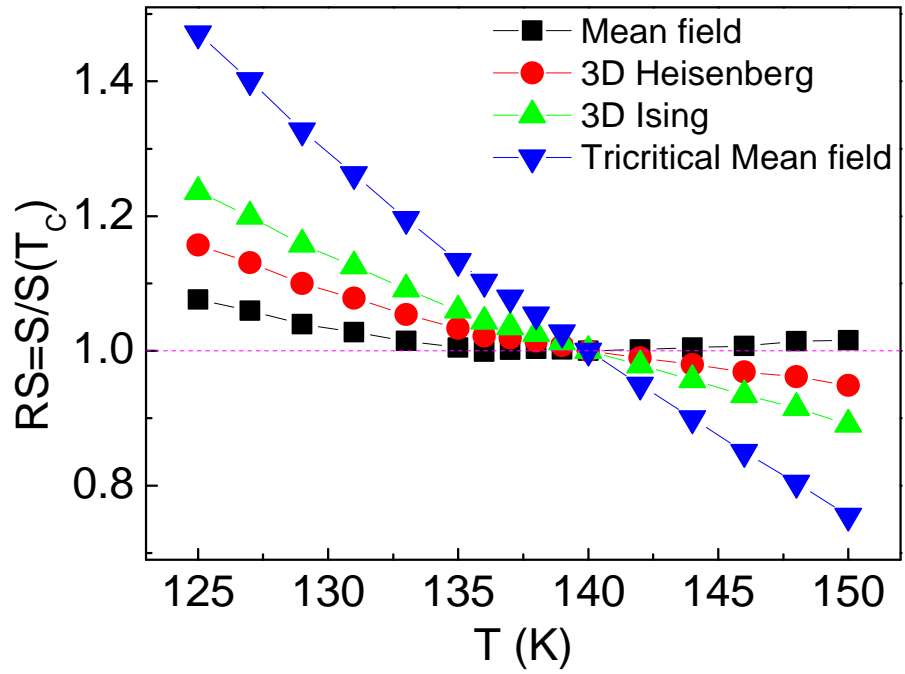

**Fig. 7.** The temperature variation of the relative slope RS [RS≡S(T)/S($T_C$)].

**Table 1:** Derived critical exponents for $Eu_{0.5}Sr_{0.5}CoO_3$ with different theoretical models and other analogous compounds reported in literatures. Abbreviations: MAP=modified Arrott plot; KFM=Kouvel-Fisher method; WSR=Widom scaling relation.

| Composition | Ref. | $T_C$ (K) | $\beta$ | $\gamma$ | $\delta$ |
|---|---|---|---|---|---|
| $Eu_{0.5}Sr_{0.5}CoO_3$ (MAP) | This work | 140.67±0.08 | 0.412±0.013 | 1.044±0.019 | 3.53(WSR) |
| $Eu_{0.5}Sr_{0.5}CoO_3$ (KFM) | This work | 140.53±0.05 | 0.415±0.002 | 1.044±0.004 | 3.51(WSR) |
| $Eu_{0.5}Sr_{0.5}CoO_3$ (M(H)) | This work | 140 | - | - | 3.06±0.03 |
| Tricritical mean-field | [56] | - | 0.25 | 1.0 | 5.0 |
| Mean-field | [39] | - | 0.5 | 1.0 | 3.0 |
| 3D-Heisenberg | [39] | - | 0.365 | 1.386 | 4.797 |
| 3D-Ising | [39] | - | 0.325 | 1.24 | 4.82 |
| $La_{0.5}Sr_{0.5}CoO_3$ | [33] | ~223 | 0.321-0.365 | 1.336-1.351 | 4.39-4.66 |
|  | [32] | 228.4 | - | 1.27±0.02 | 3.05±0.06 |